\documentclass{eas}
\usepackage{graphicx}
\begin{document}

\title{Study of the winter 2005 Antarctica polar vortex}
%
\author{F. Lascaux}\address{INAF/Osservatorio Astrofisico di Arcetri,Largo E. Fermi 5, 50125 Florence, Italy}
\author{E. Masciadri}\sameaddress{1}
\author{S. Hagelin}\sameaddress{1}
\author{J. Stoesz}\sameaddress{1}
\begin{abstract}
During winter and springtime, the flow above Antarctica at high altitude
(upper troposphere and stratosphere) is dominated by the
presence of a vortex centered above the continent.
It lasts typically from August to November.
This vortex is characterized by a strong cyclonic jet centered above the polar high.
In a recent study of our group (Hagelin \etal~\cite{hag2008}) of four different sites in the Antarctic internal
plateau (South Pole, Dome C, Dome A and Dome F), it was made the hypothesis that the wind speed strength
in the upper atmosphere should be related to the distance of the site to the center of
the Antarctic polar vortex.
This high altitude wind is very important from an astronomical point of view since it might trigger the
onset of the optical turbulence and strongly affect other optical turbulence parameters.
What we are interested in here is to localize the position of the minimum value of the wind speed at high altitude 
 in order to confirm the hypothesis of Hagelin \etal~(\cite{hag2008}).
\end{abstract}
\maketitle
\section{Introduction}
During winter, the stratospheric circulation at both poles is characterized by a large cyclonic
vortex centered in a region close to the pole (Haynes~\cite{hay2005}).
Extended climatology of the polar vortex have been already conducted, using ECMWF or NCAR-CEP data
set (Karpetchko et al. \cite{kar2005}; Waugh et al. \cite{wa1999a}) with methods based on potential 
vorticity to define the vortex.
It is well known that the vortices are much more stronger in mid-winter than in
summer (Waugh et al. \cite{wa1999b}; Harvey et al. \cite{har2002}).
In this study we want to determine the minimum speed of the high altitude wind above
the Antarctica continent. 
This might help us in identifying the center of the polar vortex and we could use this information
to qualify the sites for astronomical applications.
\par
Indeed, during a previous study about site characterization for optical turbulence
above the internal Antarctic Plateau using ECMWF analysis, our group
investigated four different sites (Hagelin et al. \cite{hag2008}): Dome A, C and F and South Pole.
One of the conclusions was that in the free atmosphere, above around 10 km, the wind speed increased
monotonically in winter proportionally to the distance to the center of the polar high.
Thus Dome C was the site showing the highest wind speed above 10 km.
We propose in this study to confirm this hypothesis looking at a climatology of the high altitude wind speed
and the corresponding vortex above the Antarctica continent for winter 2005.
Due to the weak variability of the Antarctica vortex proved in other studies, we can deduce that this one-year study
can provide a quantitative estimate with a good accuracy about the position of the minimum wind speed in altitude.
\section{The median high altitude wind speed}
We used the ECMWF analyses from MARS catalog for every day between May 1st and September 30th, 2005,
at 00 UTC.
The analysis employs the 4D-VAR assimilation scheme to assimilate a wide number of observations, including
in-situ conventional data and synthetic data from satellites.
The averaged distance between two horizontal grid points is of the order of $\sim$40~km.
To perform our study, we focused on the wind speed at two different
altitudes, 15~km and 20~km from May 1st and September 30th, every 24 hours at 00 UTC.
We computed the monthly medians of the wind speed and deduced a preferential position for the minimum of wind
speed in altitude for winter 2005.
\par
\begin{figure}
\begin{center}
\includegraphics[width=0.9\textwidth]{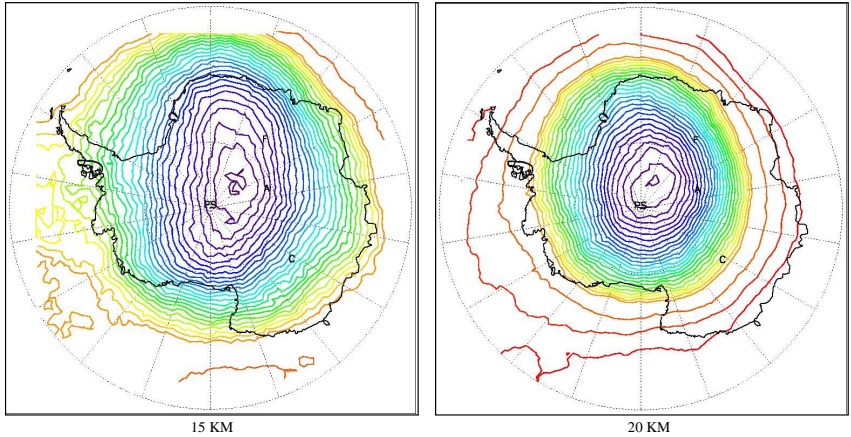}
\end{center}
\caption{Median wind speed at 15~km (column 1) and 20~km (column 2) above the Antarctic continent
         (polar stereographic projection), for Winter 2005 (May to September). Units in
         m.s$^{-1}$. Isocontours starting from 1 m.s$^{-1}$ with an increment of 1 m.s$^{-1}$.}
{\label{fig:median}}
\end{figure}

Fig.\ref{fig:median} shows the monthly median of the wind velocity at two different heights, 15 and 20~km,
computed from May 2005 to September 2005.
\par
The minimum of the median wind is clearly identified at 20~km.
The median polar vortex center (corresponding to the minimum wind speed) 
remains in a area between South Pole and Dome A. 
Between the four sites investigated by Hagelin \etal~(\cite{hag2008}), Dome C is the farthest away from the
polar vortex center, and by consequence the one with the highest wind speed in altitude.
\section{Conclusion}
This study confirms he conclusion regarding the "position space" of the polar high deduced by
Hagelin \etal~(\cite{hag2008}), and the link between the position of an Antarctic site with respect to this center
and the wind speed in altitude.
Dome C appears to have in winter a wind speed in altitude much higher than other sites like South Pole and
Dome A, closer to the center of the vortex.
Such a quantitative information should be considered by astronomers as a key issue in future plans for
astronomical facilities to be set-up in different sites of the Internal Antarctic Plateau.
\section*{Acknowledgements}
This study has been funded by the Marie Curie Excellence Grant (FOROT) - MEXT-CT-2005-023878.

\end{document}